1# Exploiting Inherent Error-Resiliency of Neuromorphic Computing to achieve Extreme Energy-Efficiency through Mixed-Signal Neurons

Baibhab Chatterjee, *Student Member, IEEE*, Priyadarshini Panda, *Student Member, IEEE*,
Shovan Maity, *Student Member, IEEE*, Ayan Biswas, *Student Member, IEEE*,
Kaushik Roy, *Fellow, IEEE* and Shreyas Sen, *Senior Member, IEEE*

*Abstract*—Neuromorphic computing, inspired by the brain, promises extreme efficiency for certain classes of learning tasks, such as classification and pattern recognition. The performance and power consumption of neuromorphic computing depends heavily on the choice of the neuron architecture. Digital neurons (Dig-N) are conventionally known to be accurate and efficient at high speed, while suffering from high leakage currents from a large number of transistors in a large design. On the other hand, analog/mixed-signal neurons are prone to noise, variability and mismatch, but can lead to extremely low-power designs. In this work, we will analyze, compare and contrast existing neuron architectures with a proposed mixed-signal neuron (MS-N) in terms of performance, power and noise, thereby demonstrating the applicability of the proposed mixed-signal neuron for achieving extreme energy-efficiency in neuromorphic computing. The proposed MS-N is implemented in 65 nm CMOS technology and exhibits > 100X better energy-efficiency across all frequencies over two traditional digital neurons synthesized in the same technology node. We also demonstrate that the inherent error-resiliency of a fully connected or even convolutional neural network (CNN) can handle the noise as well as the manufacturing non-idealities of the MS-N up to certain degrees. Notably, a system-level implementation on MNIST datasets exhibits a worst-case increase in classification error by 2.1% when the integrated noise power in the bandwidth is ~ 0.1 $\mu V^2$, along with ±3σ amount of variation and mismatch introduced in the transistor parameters for the proposed neuron with 8-bit precision.

## I. Introduction

THERE has always been a huge gap between the energy-efficiencies of the human brain and the von-Neumann model of computing which dominates the consumer market. Software simulations of the brain of a mouse with 2.5 million neurons is 9000 times slower than real-time when run on a personal computer [1]. Moreover, it consumes 400 W power as compared to the paltry 10 mW of a biological mouse brain. To emulate a human brain (100 billion

Manuscript received xxxxxxxxxx, 201x; revised xxxxxxxxxx, 201x; accepted xxxxxxxxxx, 201x. Date of publication xxxxxxxxxx, 201x; date of currentversion xxxxxxxxxx, 201x. Associate Editor: Dr. xx and Editor in Chief:Dr. xx. This work was supported in part by x.

The authors are with the School of Electrical and Computer Engineering (ECE), Purdue University, West Lafayette, IN 47907 USA (e-mail: bchatte@purdue.edu, pandap@purdue.edu, maity@purdue.edu, kaushik@purdue.edu, shreyas@purdue.edu).

Color versions of one or more of the figures in this paper are available onlineat http://xxxxxxx.

Digital Object Identifier xxxx/xxxxxxxxxx

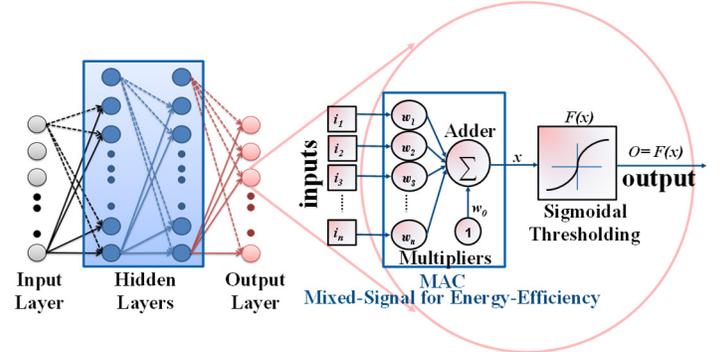

Fig. 1. Neuron model in a feedforward CNN: MAC with thresholding

neurons, 20 W power), a super computer requires 0.5 GW [2] of power. Such large differences in energy-efficiency, coupled with the rebirth of the deep learning paradigms in the last decade have forced researchers all across the world to look into alternate models of computation.

Neuromorphic computing, which loosely models the brain and uses artificial neural networks for computation, has found significant success in applications involving image and pattern recognition, miniaturized autonomous robotics [3] and neural prosthesis [4]. However, the performance and energy-efficiency of neuromorphic computing depends heavily on the choice of the neuron architecture, operating frequency, resolution and accuracy required. Digital implementation of a neuron has been the preferred choice for computing in SpiNNekar [5] and TrueNorth [6] projects due to the excellent noise immunity, variability-tolerance and technology scaling of digital designs. While SpiNNekar had no dedicated hardware for its neuron model and consumed 1 W power, IBM's TrueNorth had a dedicated point neuron model for its 1 million neurons (256 synapses each) and consumed only 65 mW. TrueNorth's primary design emphasis was on minimizing active as well as static power for a spiking neural network (SNN) by using an event-driven architecture [7], and having a compact physical design for increased parallelism on a 28 nm process that is well known for power-efficiency.

Analog/mixed-signal computational models can be easily affected by noise, variability and mismatch which makes its energy-efficiency less attractive. In an interesting study of Digital vs. Analog circuits for computing [8], the author showed that digital circuits perform better for high signal-to-noise ratio (SNR) applications (> 60 dB). However, if SNR requirements are relaxed, analog computation could be orders of magnitude more energy and area efficient. This is because analog macros, for example, a multiplier uses only one differential pair of MOSFETs which is sufficient to represent the circuit dynamics using intrinsic device parameters. On the other hand, a digital multiplier computes the same dynamics using ~1000 transistors, the combined static leakage of which could be comparable to the bias current of analog in scaled technologies. It is interesting to note that von-

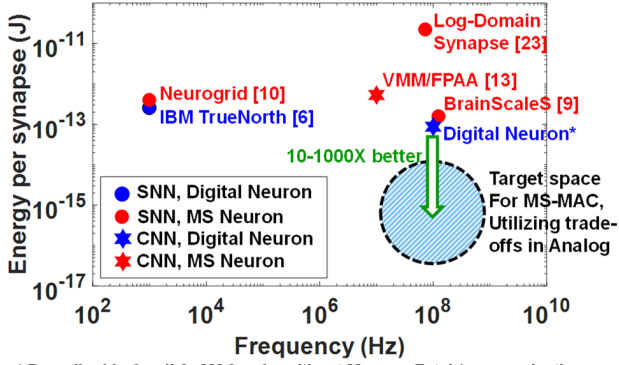

Fig. 2. Energy-Efficiency per synapse for previously reported related works, a Dig-N (Sec II A.) and target mixed-signal (MS) design

Neumann architectures depend on the high accuracy of a multi-digit representation which necessitates a digital implementation. Artificial neural networks (ANN), however, has multiple connections from inputs to output due to its distributed nature and hence the noise and variability of analog transistors can be tolerated to some extent due to this inherent error-resiliency.

*A. Related Work*

Currently, several research groups are working on the implementation of large scale SNNs with analog/mixed-signal neuron model. The BrainScales project (HICANN chip) [9] at Heidelberg University aims to develop a brain-like system that uses analog computation and digital asynchronous communication, and runs 1000–10,000 times faster than real-time. The design consists 200k analog neurons with 40 million addressable synapses, and consumes about 1kW at 125 MHz frequency. The Neurogrid project at Stanford [10] also uses a mixed design approach, and reduces transistor count further by sharing synapses and dendritic tree circuits [11]. Neurogrid has 1 million neurons, each with 8000 synapses and consumes 3.1 Watt for real-time brain computations. However, both of these designs are for SNNs which aims to model the spiking neural activities by using a current-switching neuron architecture and requires complex learning models such as spike timing dependent plasticity (STDP). Convolutional neural networks (CNN), on the other hand, can employ simple back-propagation algorithms using a multiply-and-accumulate (MAC) model [12] (shown in Fig. 1) which is more suitable than SNNs in pattern recognition applications and in scenarios involving generative-adversarial networks. In [13], a large-signal current-mode MAC implementation is demonstrated. However, in a large-signal implementation, the bandwidth of the design keeps on changing with varying bias currents, and hence the frequency of the input signals is limited by the minimum large signal current for any practical application. A small signal implementation, on the other hand, would be much more attractive in terms of the power-bandwidth trade-off. Also, a differential voltage-mode architecture would help in reducing the impact of common mode noise present in the system.

This paper presents three different architectures, leading to a compact, differential mixed-signal implementation of a MAC-based neuron that works in small-signal mode with a linear bandwidth vs. power characteristics, and can help to build an extremely energy-efficient CNN. The key contributions of this work are listed as follows:

1) This work focuses on the significant power and energy efficiency of a MS-N (over Dig-N) that can be achieved by judiciously utilizing various trade-offs of analog design at a target frequency. To the best of our knowledge, this is the first work that extensively analyzes various design trade-offs in a MS-N for energy efficiency.
2) The design of a small-signal MS-N with resistive feedback is presented, which helps to achieve much better energy-efficiency (power/bandwidth) because of a large bandwidth. Fig. 2 compares the energy efficiencies of previously reported works with the target mixed-signal design. Lower energies at the neuron level directly result in a lower power/ops at the network level for a fixed architecture and bit-precision.
3) A detailed discussion on the effects of process variability and mismatch is presented, and a method to reduce the input referred offset is demonstrated that simultaneously helps to increase the bandwidth without consuming extra power. This results in a sub-fJ MAC operation which is ~100X improvement over state-of-the-art.
4) A system level analysis of the inherent error-resiliency of neural networks is illustrated, which proves that a low to medium-complexity CNN/ANN for IoT/healthcare applications can tolerate the effects of noise and mismatch in an analog/mixed-signal design to a considerable degree.

Fig. 3. Comparison of current-mode mixed-signal MAC [13] with various voltage-mode mixed-signal MACs presented in this paper

The rest of the paper is organized as follows: Section II describes three different implementations of a neuron – a fully digital neuron, an MS-N operating in large signal mode, and an MS-N operating in small signal mode. Section III depicts the proposed MS-N with increased bandwidth and reduced offset. The specific advantages and disadvantages of these neurons are shown in Fig. 3 and will be discussed in detail in the respective sections. Section IV presents the comparison between Dig-N and the small-signal MS-N, while section V presents a detailed discussion on the trade-offs and theoretical limits of the MS-N shown in sections II-III. A system-level application of the MS-N is presented in section VI, wherein the error-resiliency of a CNN and a fully connected network is demonstrated separately. Section VII compares our design with other state-of-the-art neuron architectures. We conclude the work in Section VIII by summarizing our major contributions.

II. NEURON ARCHITECTURES: DIG-N AND MS-N

It is well-established that analog design is superior to digital in terms of power and area for applications that require < 8-bit precisions [8]. It is also indicated in [14] that anything > 8-bit fixed point precision is redundant for most ANN applications. In this work, our target application is a classification problem for digit/image recognition, using the MNIST dataset [15] for handwritten digits and the CIFAR-10 dataset [16] for images. The convolutional neural network (CNN) and fully connected neural network (FCN) architectures are used at the system level, as will be shown in Section VI. Fig. 4 shows the classification error for the applications and different network architectures as a function of the fixed-point bit-precision. It can be observed that the classification error does not increase significantly from the baseline if the precision is reduced from 16-bit to 6-bit. The error rate starts to increase significantly at 3-bit precision. 2-bit precision results in a classification error > 80%. Based on these numbers, we have limited ourselves with digital and mixed-signal neurons having a precision in the range of 3-8 bits.

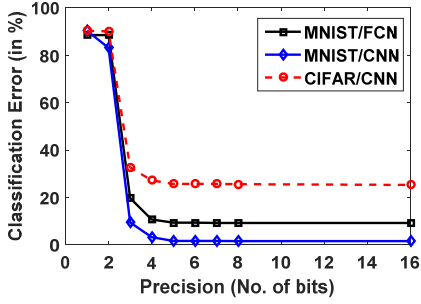

Fig. 4. System-level Precision analysis for target applications

*A. Digital Neuron*

The basic functionality of a MAC-based neuron is to evaluate the weighted sum of input signals followed by a thresholding function for activation. Hence, based on the nomenclature used in Fig. 1, we can write the output of a neuron as $o = F(\Sigma x_i w_i)$ for $i = 1,2,3,...,n$ where $n$ is the total number of synapses, $F$ is the activation function, which can be hard-limiting (e.g. step function) or soft-limiting (e.g. log/tan-sigmoid or rectified linear function). $w_i$ is the weight corresponding to the $i$-th multiplier having input voltage $x_i$. The number of bits ($N$) in $w_i$ or $x_i$ defines the precision of the MAC architecture. An 8-bit MAC needs an 8×8-bit multiplier and 17-bit adder, while a 3-bit MAC needs a 3×3-bit multiplier and 7-bit adder.

We have synthesized an 8×8-bit Wallace tree (WT) multiplier with a 17-bit ripple-carry adder in 65 nm CMOS technology, along with comparators for activation logic. The 3-bit version of the same design uses a 3×3-bit Wallace tree multiplier and a 7-bit ripple carry adder. A carry look-ahead adder or a carry-save adder does not result in significant speed advantage at such low precisions at the expense of more hardware. Although WT multipliers are fast, they consume higher power than most other multiplier architectures. For this reason, we have also synthesized 8×8-bit and 3×3-bit array multipliers (AM) which consume less power. The number of different cells and transistors in an N-bit digital MAC (both WT and AM) is given in Table I. Unlike analog implementations that rely on intrinsic device dynamics, digital logic computes the circuit dynamics algorithmically, and this makes the number of transistors as high as 2500 for the 8-bit case.

*1) Bandwidth*

Bandwidth of the Dig-N is dependent on the supply voltage ($V_{DD}$), as shown in Fig. 5. The BW is slightly higher for the 3-bit Dig-N as the critical-path delays are less for a low-complexity design.

*2) Power vs. Performance*

The power in digital circuits usually consists two components, a) dynamic power and b) static leakage power. Dynamic power in the design is given by Eq. 1.

$$P_{Dig} = \sum_{i=1}^{N_{Dig}} \alpha_i C_i V_{DD,Dig}^2 f \quad (1)$$

Where $\alpha_i$ is the activity-factor of the $i$-th node, $N_{Dig}$ is the total number of nodes, $C_i$ is the capacitance at a switching node, $V_{DD,Dig}$ is the supply voltage and $f$ is the operating frequency.

The static leakage current is due to a) subthreshold conduction, b) reverse-biased p-n junction conduction and c) gate induced drain leakage (GIDL), out of which subthreshold conduction is the dominant factor [17]. Fig. 6 shows total power vs. frequency for the 8-bit and a 3-bit Dig-Ns designed in 65nm CMOS process. The dynamic power dominates for frequencies > 10 MHz. However, at lower frequencies, power consumption is dominated by leakage, which increases proportional to the number of transistors (8-bit WT has ~13 times more leakage than that of a 3-bit AM Dig-N, which corresponds to the ratio of transistors present in corresponding designs). The minimum energy

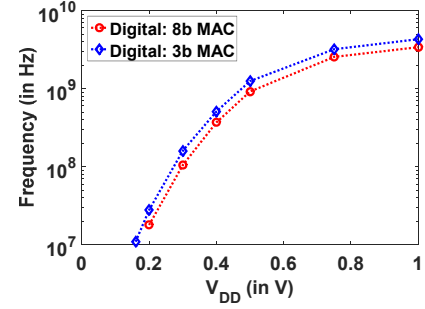

Fig. 5. Bandwidth of the Dig-N (WT) with different supply voltages

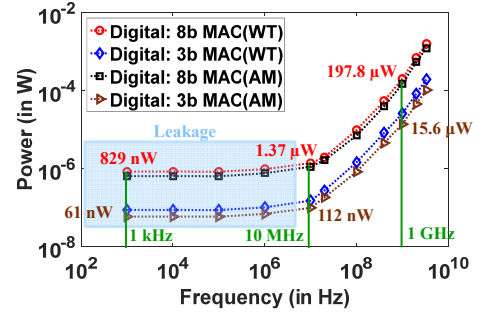

Fig. 6. Power consumption of the Dig-Ns across frequencies

TABLE I
NUMBER OF CELLS AND TRANSISTORS IN *N*-BIT DIGITAL MAC

| N | Number of AND gates | Number of (HA, FA)[a] | Total Number of transistors |
|---|---|---|---|
| 3 | 9  | (2,8)-WT[c], (3,3)-AM[d] | 426-WT, 234-AM |
| 4 | 16 | (3,14)-WT, (4,8)-AM      | 738-WT, 504-AM |
| 5 | 25 | (4,22)-WT, (5,15)-AM     | 1146-WT, 870-AM |
| 6 | 36 | (9,30)-WT, (6,24)-AM     | 1638-WT, 1332-AM |
| 7 | 49 | (12,42)-WT, (7,35)-AM    | 2274-WT, 1890-AM |
| 8 | 64 | (14,56)-WT, (8,48)-AM    | 2988-WT, 2544-AM |

[a]HA: half-adder, FA: full-adder
[c]WT: Wallace-tree based MAC, [d]AM: Array-Multiplier based MAC

consumption of the 8-bit AM Dig-N can be calculated as 87 fJ@10 MHz (137 fJ for WT, 8-bit).

*3) Noise*

The dominant source of noise in digital circuits is quantization noise. Thermal noise induced bit-flipping is practically rare due to high noise margin. Assuming a uniform distribution of error, the quantization noise voltage (in Volts) of the Dig-N can be expressed as shown in Eq. 3.

$$N_Q = \frac{(V_{high} - V_{low})}{\sqrt{12(2^N - 1)}} \quad (3)$$

Where $N$ is the number of bits (precision), and $(V_{high} - V_{low}) \approx V_{DD}$. As $N$ is reduced, $N_Q$ increases exponentially. For *N*-bit precision, the SNR in presence of $N_Q$ can be calculated as $SNR = 6.02N + 1.76$ (in dB). This results in 50 dB SNR for 8-bit precision, and 20 dB SNR for 3-bit precision.

*B. Mixed-Signal Neuron (large-signal mode)*

The Dig-N is not energy efficient at frequencies < 10 MHz where it suffers from static leakage power. Mixed-signal neurons (MS-N) with analog computation can potentially have far better energy efficiency, as they can be designed with only a few transistors unlike Dig-Ns. Fig. 7 shows an *N*-bit, differential-amplifier based sub-threshold MS-N architecture with $n$ synaptic weights. The *N*-bit weights are coming from a digital memory while the MAC operation is performed in an analog fashion, hence the name MS-N. Bit *j* (*j* = 0,1,2, … , *N-1*) of the *i*-th weight activates switches at the tail current sources for each of the $2^j$ slices (from slice number ($2^j$-1) to slice number ($2^{(j+1)}$-2)) in the *i*-th multiplier (for all *i* = 1,2,3, … ,*n*). The tail current source for each slice has a value of $I_{unit}$ when on. Thus the total bias current through the load keeps on changing with the weight, which means the effective bandwidth of the system corresponds to the minimum weight while the power consumption of the system corresponds to the maximum weight. Moreover, changing the large signal current with the weight necessitates a resistive load to ensure linearity of multiplication (a PMOS load will be non-linear). This leads to significant area penalty for an on-chip implementation.

Fig. 8 illustrates the power consumption in the MS-N with respect to frequency, and compares it to the power consumption of the digital neurons. The 8-bit analog MAC has a constant energy consumption of $\approx$ 0.9 pJ across all frequencies, and has better power-efficiency than digital MACs at frequencies < 1 MHz. The 3-bit analog MAC has better power efficiency than the 3-bit digital MACs at all frequencies.

*C. Mixed-Signal Neuron (small-signal mode)*

To achieve a better power-bandwidth trade-off, the weights can be used to activate switches at the gate of the input sub-threshold transistors while a fixed current $I_{bias} = \sum_{j=1}^{2^{(N-1)}} jI_{unit} = (2^N - 1)I_{unit}$ flows through the *i*-th multiplier (for all *i* = 1, 2, 3, … ,*n*), enabling a small signal mode of operation. When a switch is off, the corresponding input is connected to ground to avoid floating nodes. A PMOS load can now be employed as the bias current through the neuron is fixed. Since the gain of the neuron needs to be ≤ 1 (depending on the weight) to avoid saturating subsequent stages, high-impedance PMOS loads now allow a smaller effective transconductance which leads to better energy-efficiency. Also, the number of slices in the *i*-th multiplier is now reduced to *N* (from ($2^N$-1) in the large signal case) while the current source of the *j*-th slice now carries a current of $j \times I_{unit}$. This reduces the effective capacitance at the output of the neuron, thus increasing its bandwidth. This configuration is illustrated in Fig. 9. The output of the MS-N is modeled in Eq. 4.

$$V_{out} = F(A_v \times \sum_{k=1}^{n} w_k v_k) \quad (4)$$

Where *F* is the voltage transfer function of a differential amplifier, which acts as a sigmoidal activation function in the context of a neuron, $w_k$ and $v_k$ are the weights and the ac-coupled input voltage of the *k*-th multiplier, respectively. $A_v$ is the small-signal voltage gain of a multiplier where $A_v \approx k \times g_{m_{mul}}/g_{m_p}$, $g_{m_{mul}} = \sum_{j=1}^{N} g_{m_{in,j}}$ is the input transconductance of a multiplier ($g_{m_{in,j}}$ is the input transconductance in the *j*-th slice of a multiplier), $g_{m_p}$ is the transconductance of the PMOS load, and $(g_{m_{mul}} \sum_{k=1}^{n} w_k v_k)$ represent the KCL addition of currents at the output nodes. The detailed expression for $A_v$ is shown in Eq. 5.

$$A_v = -\frac{C_{coupling}}{C_{coupling} + N \times C_{gg}} \times \frac{\sum_{j=1}^{N} g_{m_{in,j}}}{g_{m_p} + g_{ds_p} + \sum_{j=1}^{N} g_{ds_{in,j}}} \quad (5)$$

Where $C_{coupling}$ is the AC coupling capacitance for each multiplier (100 fF – not shown in Fig. 9), $C_{gg}$ is the effective gate to ground capacitance at the input of a multiplier, $g_{m_{in,j}}$ and $g_{ds_{in,j}}$ are the transconductance and output conductance, respectively for the input NMOS transistors in the *j*-th slice of a multiplier. Similarly, $g_{m_p}$ and $g_{ds_p}$ are the transconductance and output conductance, respectively, for the PMOS load. The $g_m$ quantities are in the same range of $g_{ds}$, and hence $g_{ds}$ cannot be ignored. Any source resistance/resistance at gate is ignored as it will be very small and will create a non-dominant pole at a very high frequency.

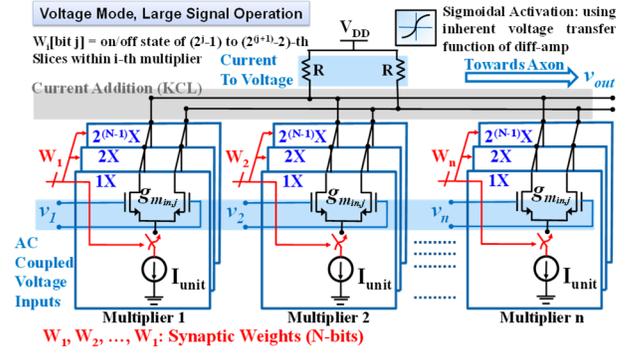

Fig. 7. A MAC-based MS-N operating in large-signal mode: Total number of slices in *k*-th Multiplier = $1 + 2 + \cdots + 2^{(N-1)} = 2^N$

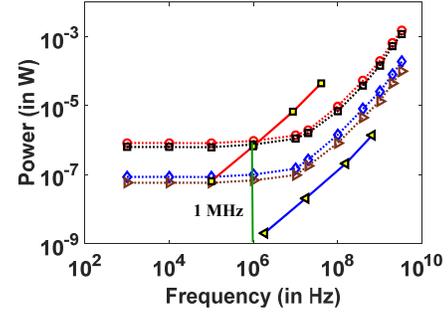

Fig. 8. Power consumption of the large-signal MS-MAC vs. frequency

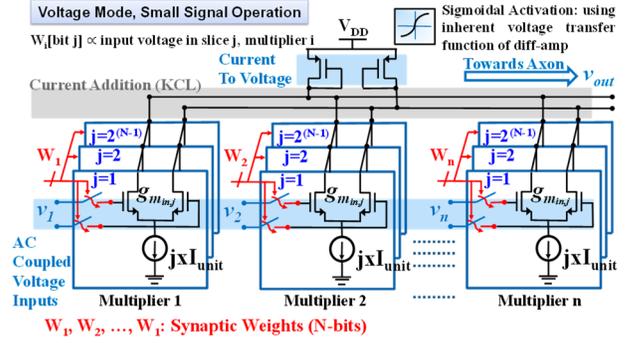

Fig. 9. A MAC-based MS-N operating in small-signal mode: Total number of slices in in *k*-th Multiplier = *N*

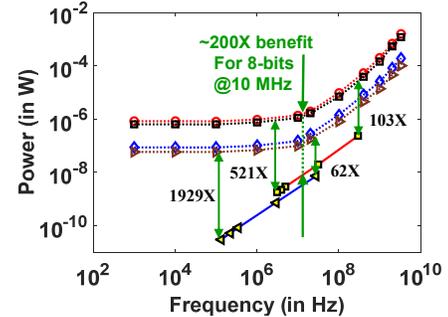

Fig. 10. Power consumption of the small-signal MS-MAC vs. frequency

*1) Bandwidth*

The linearity of bandwidth w.r.t. power is substantiated from Eq. 6 which shows the bandwidth at the output node of an $N$-bit synapse in terms of the subthreshold current.

$$BW = \frac{g_{m_p}}{2\pi \times C_{eff}} = \frac{(2^N-1)I_{unit}}{4\pi \times \eta V_T \times C_{eff}} \quad (6)$$

Where $g_{m_p} = \frac{I_p}{\eta V_T}$ is the transconductance of the PMOS load, $I_p$ is the large-signal current through the PMOS load which is equal to half of the total current through a multiplier, $C_{eff}$ is the effective output capacitance, and $I_{unit}$ is the unit current of a multiplier. All input transistors are designed to carry sub-threshold current.

*2) Power vs. Performance*

The total current is close to 255 times of the unit current for 8-bit MS-N and 7 times of the unit current for 3-bit MS-N. Fig. 10 demonstrates the power consumption in the MS-N with respect to frequency. The linearity of frequency vs. power consumption is established from this figure as well. The energy consumption is constant (0.8 fJ for the 8-bit MS-N at all frequencies) and is > 60X better than the AM-based Dig-N.

*3) Noise*

Unlike Digital-MAC, MS-MAC does not have a noise margin; hence the accuracy of the network will be affected by the noise in the circuit. The main components of noise in the analog MAC are:

a) MOSFET thermal noise: This is the dominant analog noise source. Considering the input and load transistors are in sub-threshold saturation for each multiplier, we calculate the open circuit mean-square noise power per unit bandwidth. The total thermal noise current power per unit bandwidth for the subthreshold input transistors connected to each polarity of the differential input in a multiplier is given by Eq. 7 [18].

$$\overline{i_{n,in}^2} = 2qI_n \quad (7)$$

Where $q$ is electronic charge and $I_n$ is the total bias current through the relevant input transistors. Interestingly, $I_n$ is half of the total current through each multiplier, and hence Eq. 7 can be written as Eq. 8 for an N-bit synapse.

$$\overline{i_{n,in}^2} = 2q \times \tfrac{1}{2}\sum_{i=1}^{N} 2^{(i-1)} \times I_{unit} = q \times (2^N - 1)I_{unit} \quad (8)$$

The channel noise for the PMOS load in sub-threshold is given by Eq. 9.

$$\overline{i_{n,p}^2} = 2qI_p = 2q \times \tfrac{1}{2}(2^N - 1)I_{unit} = q \times (2^N - 1)I_{unit} \quad (9)$$

Hence, the total thermal noise power at the output (in $V^2$/Hz) can be obtained as given in Eq. 10.

$$\overline{v_n^2} = [q \times (2^N-1)I_{unit} + q \times (2^N-1)I_{unit}] \times \left(\frac{1}{g_{m_p}}\right)^2$$
$$= [4qI_p] \times \left(\frac{\eta V_T}{I_p}\right)^2 = \frac{4q\eta^2 V_T^2}{I_p} = \frac{8q\eta^2 V_T^2}{(2^N-1)I_{unit}} \quad (10)$$

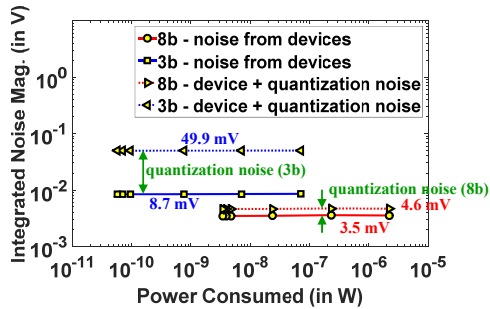

Fig. 11. Integrated device and quant. noise vs. power for the MS-MAC

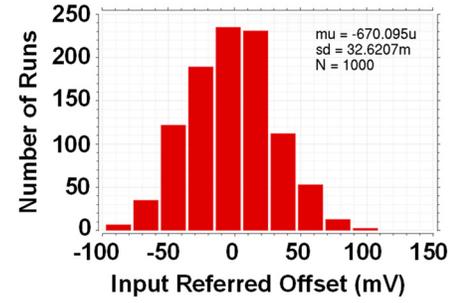

Fig. 12. Mismatch Analysis: input referred offset for the 8-bit MS-MAC

Therefore, $\overline{v_n^2}$ is inversely proportional with bias current. Eq. 6 and Eq. 10 also imply that the integrated thermal noise power over the bandwidth will be constant, given by Eq. 11.

$$\overline{v_{n,integrated}^2} = \frac{4q\eta V_T}{2\pi \times C_{eff}} \quad (11)$$

This means that to reduce the integrated thermal noise, we need to use larger $W$ and $L$ for the transistors which increases $C_{eff}$. However, this also reduces bandwidth. We will show a method to overcome this trade-off in section III.

b) Flicker noise: This is a ubiquitous noise present in all electronic systems, which is most significant at low frequencies. This noise power is empirically given by Eq. 12.

$$\overline{v_{n,f}^2} = \frac{K}{f^\alpha} \quad (12)$$

where $K$ is an empirical parameter which is dependent on device type, dimensions and technology node, and $\alpha$ is an exponent which is usually close to unity [19].

c) Switch noise: Since we have introduced switches in the signal path in our design, any switching activity will give rise to transient noise. However, it must be noted that the weights will be set during training and hence there would be no noise from the switches during the testing phase.

d) Quantization noise: A binary code is used to activate the binary weights that connect the inputs to the desired differential pairs. Thus, effectively, a DAC operation is being carried out, which gives rise to quantization noise, given by Eq. 3. This is found to be the overall dominant noise, because of the low precisions in our application.

In the system level applications, however, the integrated thermal and flicker noise is of more importance, as quantization noise affects bit precision analysis for the weights (Fig. 4) and creates the same baseline error for both Dig-N and MS-N. Fig. 11 exhibits the noise power (in $V^2$) integrated over the signal bandwidth, as a function of the total power consumed. As expected, this is relatively constant since bandwidth and noise floor (in $V^2$/Hz) both scales linearly in equal and opposite amounts with bias current.

*4) Effect of mismatch/DC offset*

There is negligible systematic offset because of the symmetry of the design. However, there will be a random offset because of mismatches in threshold voltages and dimensions during fabrication. These mismatches can be within-die (local) or die-to-die (global) and creates an offset at the output nodes of each multiplier. Since the individual multipliers are AC coupled (with 100 fF coupling capacitance), this offset will not propagate to subsequent stages. However, the input bias points of the two legs of a branch can be different due to the offset, causing variations in gain and swing in the two branches.

The primary contributor to overall mismatch is often the mismatch in threshold voltages ($V_{TH}$) of the otherwise symmetric transistors [22]. The standard deviation for $V_{TH}$ is given by Eq. 13.

$$\sigma_{V_{TH}} = \frac{A_{V_{TH}}}{\sqrt{WL}} + S_{V_{TH}} \times D_x \quad (13)$$

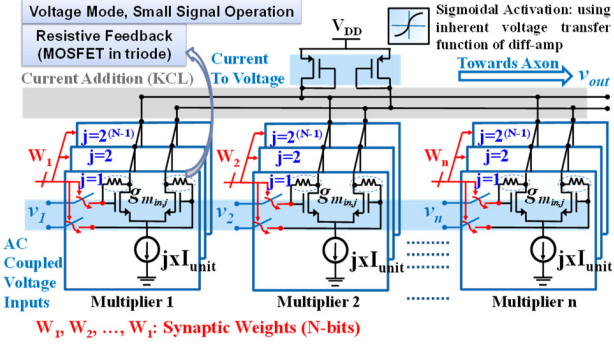

Fig. 13. Proposed small-signal MS-N with resistive feedback: : Total number of slices in in $k$-th Multiplier = $N$

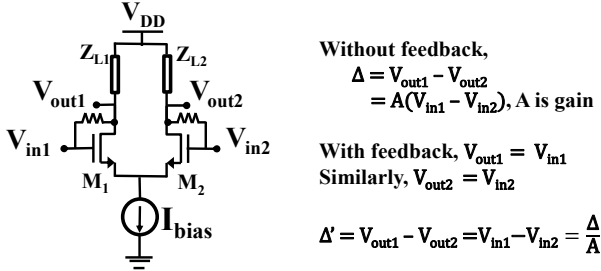

Fig. 14. Offset Compensation using resistive feedback: all voltages are DC

Where $A_{V_{TH}}$ is the $V_{TH}$-mismatch parameter, $W$ and $L$ are device dimensions, $D_x$ is the distance between the centers of the devices, and $S_{V_{TH}}$ is the distance proportionality constant ($\approx 0$ for common centroid layout). Mismatches in sizing, mobility, gate oxide capacitance and body bias parameter also contributes to the overall offset. With the standard values of the parameters for a 65 nm process and the values of $W$ and $L$, $3\sigma_{V_{TH}}$ is calculated to be around 60 mV for each differential leg in our design. Fig. 12 shows the results of the Monte-Carlo analysis which indicates a $3\sigma$ output offset variation of 98 mV, which is alarmingly high considering the voltage swing to be a few hundred mV.

### III. PROPOSED MIXED SIGNAL NEURON: REDUCED MISMATCH/OFFSET AND INCREASED BANDWIDTH

To solve the issue of offset, we propose a mixed-signal neuron with resistive feedback, as shown in Fig. 13. The feedback resistance tries to keep the input and output DC points at the same voltage. At the same time, the resistance forms a low-pass filter with the parasitic $C_{gd}$ in the feedback path. This creates a zero in the feed-forward transfer function which, when superimposed on the dominant pole, increases the bandwidth of the circuit. The increase in bandwidth enables us to increase the $W$ and $L$ of the MOSFETs (keeping the ratio same) that reduces the input offset and the effect of mismatch/noise. These benefits are obtained at no extra power cost, and a minimal area cost.

Since on-chip poly-resistors consume a significant area, the resistance in the feedback path is implemented using an *NMOS in triode*. The non-linearity introduced due to this does not affect the final results since the input common mode range of the neuron is observed to be > 150 mV across process corners.

The gain of this structure is given by Eq. 14.

$$A_v = -\frac{C_{coupling}}{C_{coupling}+N \times C_{gg}} \times \frac{\sum_{j=1}^{N} g_{m_{in,j}} - \frac{N}{R}}{g_{m_p}+g_{ds_p}+\sum_{j=1}^{N} g_{m_{in,j}}+\frac{N}{R}} \quad (14)$$

Where $R$ is the feedback resistance (an NMOS in triode). Again, $g_m$ quantities are in the same range of $g_{ds}$, and hence $g_{ds}$ terms cannot be ignored. The dominant pole is given by Eq. 15, while Eq. 16 models the zero.

$$\omega_{p,dom} = \frac{g_{m_p}+\frac{N}{R}}{\sum_{j=1}^{N} C_{gd_j}} \quad (15)$$

$$\omega_z = \frac{\sum_{j=1}^{N} g_{m_{in,j}} - \frac{N}{R}}{\sum_{j=1}^{N} C_{gd_j}} \quad (16)$$

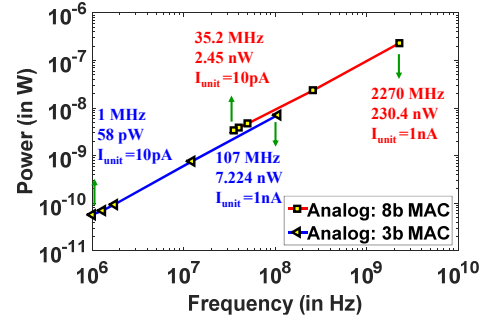

Fig. 15. Power consumption of the proposed MS-MAC vs. frequencies

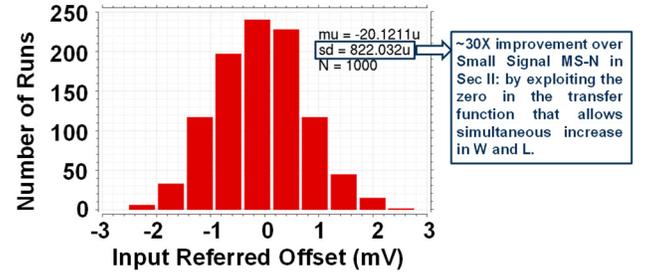

Fig. 16. Mismatch Analysis: input referred offset for proposed MS-MAC

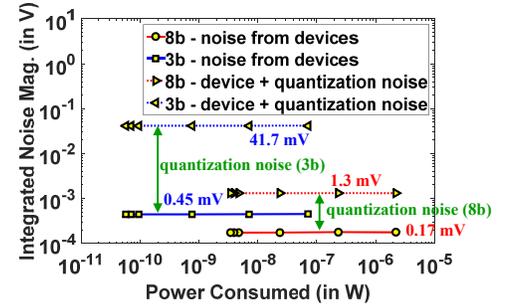

Fig. 17. Integrated noise vs. power for the proposed MS-MAC

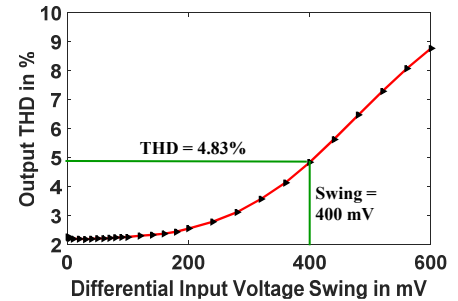

Fig. 18. THD for the 8-bit proposed MS-MAC, with $I_{unit}$ = 100 pA

Solving for $\omega_{p,dom} = \omega_z$, we can find $R$, as given by Eq. 17.

$$R = \frac{2N}{\sum_{j=1}^{N} g_{m_{in,j}} - g_{m_p}} \quad (17)$$

Fig. 13 shows how offset compensation can be achieved using the resistive feedback structure. The output offset with the feedback is reduced by a factor of A as compared to the offset without the feedback. In our design, A=1 which results in a residual offset same as the original offset, but the introduction of the zero due to the resistive feedback enables bandwidth extension which allows larger devices (hence, smaller offset). Fig. 15 exhibits the frequency (bandwidth) vs. power characteristics of the proposed MS-N. The energy efficiency is ~ 100 aJ for the 8-bit design, which is much lower than the Dig-N (87 fJ, best case for AM-based MAC) and the other designs of the MS-Ns presented in previous sections (0.8 fJ, best case). Fig. 16 exhibits a $3\sigma$ output offset variation of 2.5 mV from Monte-Carlo simulations of the 8-bit design, while Fig. 17 shows that the integrated output noise power < 0.1 μV$^2$ over the bandwidth. Thus the overall worst case effect of noise and mismatch can be considered to be within 3 mV. Fig. 18 presents the output total harmonic distortion (THD) as a function of the voltage swing when $I_{unit} = 100$ pA. With a differential input swing of 400 mV, THD is < 5%. Hence the 3 mV error due to mismatch and noise is within 1% of the output swing.

*1) Effect of Variability*
Apart from noise and mismatch, the MS-N also suffers from process variations. Table II lists the important specifications of the MS-N across different process corners. Since the gain is always close to 1 and the integrated noise is almost constant across process corners, the only limitation posed by the process corners is the bandwidth when NMOS transistors are slow. However, even the worst case bandwidth results

TABLE II
PROCESS VARIATION IN PROPOSED 8-BIT MS-N[1]

| Specification | TT Process | FF Process | SS Process | FNSP Process | SNFP Process |
|---|---|---|---|---|---|
| Power (nW) | 20.8 | 24.49 | 19.32 | 23.44 | 20.27 |
| Bandwidth (MHz) | 292.2 | 374 | 247.4 | 410.3 | 231.7 |
| Gain (dB) | -0.457 | -0.361 | -0.696 | -0.315 | -1.39 |
| Integrated Noise Power (V$^2$) | 6.3e-8 | 5.4e-8 | 6.8e-8 | 6.4e-8 | 6.6e-8 |
| Energy Efficiency (Power/Bandwidth) | 71 aJ | 65 fJ | 78 aJ | 57 aJ | 87 aJ |

[1]: Results are with $I_{unit} = 100$ pA

in an energy efficiency that is much better than existing architectures, as will be seen in Section VII.

*2) Stability*
The bandwidth is extended by making $\omega_{p,dom} \approx \omega_z$. The feedback resistor can lead to oscillations if the phase margin is not enough. However, the system is always stable because the gain of the system is ≤ 1, and hence the concept of phase margin does not apply (as there is no gain crossover frequency).

*3) DNL and INL*
The multiplication of the input signal with the weights is effectively a DAC operation. The slices in each multiplier are designed with binary weighted bias currents but with same sized input transistors, and hence the overdrives are different for the input transistors in each slice, which leads to an effective transconductance that increases in a slightly non-linear manner with the weight. Hence, this architecture achieves high bandwidth and low power at the cost of nonlinearities in the DAC operation. However, with large input swing (~400 mV differential) and a small unit current (~100 pA), the differential non-linearity (DNL) and integral non-linearity (INL) can be kept within ±0.5 LSB, even in the presence of $\pm 3\sigma$ amount of mismatch as shown in Fig. 19 and Fig. 20. This implies that there is no missing code during the MS-N operation.

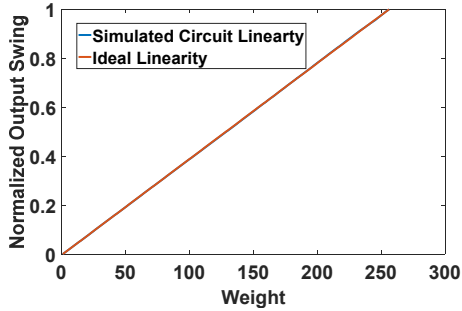

Fig. 19. Linearity for the 8-bit MS-N with 400 mV differential swing and 100 pA unit current (in presence of $\pm 3\sigma$ mismatch in device dimensions)

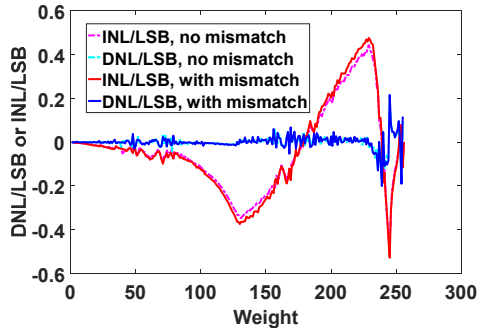

Fig. 20. Simulated INL and DNL of the MS-N (conditions same as Fig. 19)

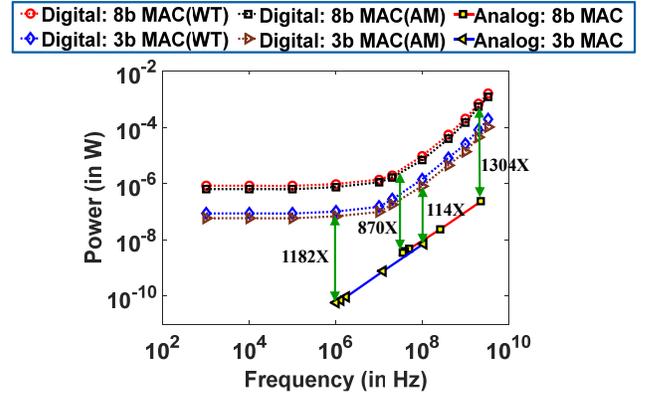

Fig. 21. Power consumption of the proposed MS-MAC against Dig-MAC

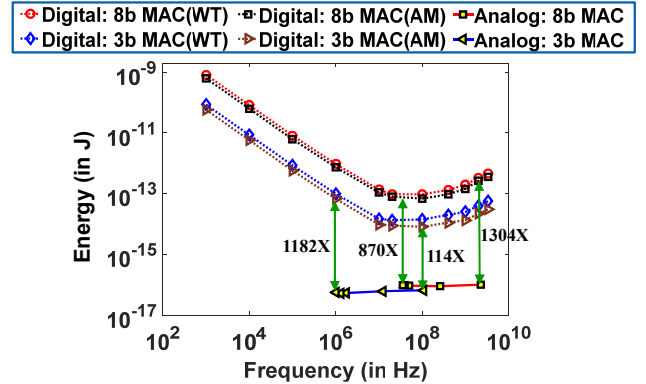

Fig. 22. Energy consumption of the proposed MS-MAC against Dig-MAC

## IV. COMPARISON: DIG-N VS. PROPOSED MS-N

The power and energy consumption of the proposed MS-N at different frequencies is shown in Fig. 21 and Fig. 22, respectively. The proposed MS-N is 2-3 orders more energy- efficient as compared to the Dig-N over all frequencies. At frequencies < 1 MHz, the all-digital implementation suffers from static leakage currents, which is quite high due to the large number of transistors. Energy consumption at frequencies >500 MHz is also high for the Dig-N because of dynamic voltage and frequency scaling (DVFS). In contrast, power for the analog MAC in the MS-N scales linearly with frequency and is more energy-efficient both at low and high frequencies. For the 8-bit design, the MS-N is ~870X better than Dig-N in terms of energy-efficiency at near-threshold point (~10 MHz), which is a further 4X improvement over the performance of the MS-N presented in section II. C and in [24].

## V. THEORETICAL LIMITS AND TRADE-OFFS FOR MS-N

To understand the performance benefits of the MS-N over Dig-N, we consider the fundamental design of the 8-bit MS-N presented in section II. C. The power consumption for a MS-multiplier ($P_{MS}$) in that case is calculated as given in Eq. 18.

$$P_{MS} = V_{DD,Ana}I = V_{DD,Ana} \times (2\pi \times \eta V_T \times C_{eff} \times BW) \quad (18)$$

Where $V_{DD,Ana}$ is the supply voltage for the analog MAC, $I=(2^N - 1)I_{unit} = (2\pi \times \eta V_T \times C_{eff} \times BW)$ is the total current in the multiplier, as found from Eq. 6. On the other hand, The dynamic power consumption of a Dig-N ($P_{Dig}$) was shown in Eq. 1, which leads us to Eq. 19, which shows the ratio $\left(\frac{P_{Dig}}{P_{MS}}\right)$.

$$\frac{P_{Dig}}{P_{MS}} = \frac{\sum_{i=1}^{N_{Dig}} \alpha_i C_i V_{DD,Dig}^2 f}{V_{DD,Ana} \times (2\pi \times \eta V_T \times C_{eff}) \times BW} \quad (19)$$

For iso-frequencies ($f$ = BW) and considering $V_{DD,Ana}$ = 1V, $V_{DD,Dig}$ = 0.4 V (which are the values of analog and digital supplies in the simulations), $\eta$ = 1.3 and $V_T$ = 0.026 V, we get Eq. 20.

$$\frac{P_{Dig}}{P_{MS}} = \frac{V_{DD,Dig}^2}{V_{DD,Ana} \times (2\pi \times \eta V_T)} \times \frac{\sum_{i=1}^{N_{Dig}} \alpha_i C_i}{C_{eff}} = 0.75 \times \frac{\sum_{i=1}^{N_{Dig}} \alpha_i C_i}{C_{eff}} \quad (20)$$

Depending on whether the effective output capacitance of the MS-multiplier ($C_{eff}$) is dominated by it's own intrinsic capacitance (from the slices), or by the number of multipliers connected at the output node, or by the number of fan-outs from the neuron, the following scenarios may arise:

*1) Case 1: Intrinsic slice capacitance dominates*

From the simulation results in Fig. 6 and Fig. 10, it is clear that the worst case energy-benefit $\left(\frac{P_{Dig}}{P_{MS}}\right)$ for the MS-N would be observed near 10 MHz frequency where the energy consumption of Dig-N is at its lowest. At that frequency, $\sum_{i=1}^{N_{Dig}} \alpha_i C_i$ for the Dig-N can be approximated as $\frac{(1370-829)\times 10^{-9}}{0.4 \times 0.4 \times 10^7} \approx 338$ fF. Hence,

$$\frac{P_{Dig}}{P_{MS}} = \frac{0.75 \times 338}{C_{eff}} = \frac{253.5}{C_{eff}} \quad (21)$$

If all the slices in the MS-N were designed with unit current, then the unit slice would have required to be repeated $2^j$-times for the $j$-th bit of the weight for a multiplier (as shown in Sec. II B., $j = N - 1$ is the MSB, and $j = 0$ is the LSB). This would have meant $C_{eff} = (2^N - 1)C_{unit} = 255 \times C_{unit}$ where $C_{unit}$ is the unit capacitance from each slice, connected to the output node. As a result, $\frac{P_{Dig}}{P_{MS}}$ would have been $\frac{253.5}{255 \times C_{unit}}$, which is close to a mere factor of 2.5 assuming $C_{unit} = 0.4$ fF which is the unit node capacitance in both our Dig-N and MS-N implementations. Hence, to get the energy-benefits of MS-N, irregular slices had to be adopted as shown in Fig. 9, where the tail current sources are binary-weighted, but each bit of the weight is connected to only one slice. This implies that $C_{eff} = NC_{unit} = 8 \times C_{unit}$ and hence $\frac{P_{Dig}}{P_{MS}} = \frac{253.5}{8 \times 0.4} \approx 79$. Of course, this energy benefit comes with the issue of non-linearity as discussed in section III.

The analysis shown above only considers the dynamic power of the Dig-N. The total energy benefit considering dynamic + leakage current for each node in the Dig-N in this scenario is given by Eq. 22 (again taking the numbers from Fig. 6).

$$\frac{P_{Dig}}{P_{MS}} = \frac{1370 \times 10^{-9}}{V_{DD,Ana} \times (2\pi \times \eta V_T) \times C_{eff} \times 10^7} = \frac{630.6}{C_{eff}} \quad (22)$$

Which results in $\frac{P_{Dig}}{P_{MS}} \approx 197$ when $C_{eff} = 8 \times C_{unit}$ and $C_{unit} = 0.4$ fF. These results correspond to the graph shown in Fig. 10 where the energy benefit for the MS-Multiplier is about 200X that of the digital implementation near 10 MHz. Theoretically, we can arrive at the same result from Eq. 20 by noting that $\frac{\sum_{i=1}^{N_{Dig}} \alpha_i C_i}{C_{eff}}$ is essentially the ratio of number of transistors in Dig-N ($N_{Dig}$) to the number of input transistors in MS-N ($N_{Ana}$), multiplied with the effective $\alpha$ of the Dig-N. Hence, $\frac{P_{Dig}}{P_{MS}}$ can be expressed as Eq. 23.

$$\frac{P_{Dig}}{P_{MS}} = 0.75 \times \alpha \times \frac{N_{Dig}}{N_{Ana}} \times L \quad (23)$$

Where $L$ is the additional contribution from leakage ($L \approx 2.5$ near the threshold frequency of 10 MHz where dynamic power is almost 67% of the leakage power). Writing $N_{Ana} = \frac{N_{Ana,regular}}{\left((2^N-1)/N\right)}$ ($N$ is the number of bits), we get Eq. 24.

$$\frac{P_{Dig}}{P_{MS}} = 0.75 \times \alpha \times \frac{N_{Dig}}{N_{Ana,regular}} \times \frac{2^N - 1}{N} \times L \quad (24)$$

Activity Factor (Dig-N) | Ratio of Transistors | Benefit from Irregular slices | Leakage (Dig-N)

$$= 0.75 \times 0.3 \times \frac{2805}{255} \times \frac{2^8 - 1}{8} \times 2.5$$

$$= 197$$

Where $N_{Ana,regular}$ is the number of input transistors in MS-N with regular slices, and $\frac{2^N - 1}{N}$ denotes the benefit gained from irregular slices (as the number of transistors reduce from $(2^N - 1)$ to $N$). In the calculation, we have assumed $N_{Dig} \approx 2805$ (from Table I) and $\alpha \approx 0.3$ ($\sum_{i=1}^{N_{Dig}} \alpha_i C_i = 338$ fF, hence $\alpha$ can be calculated and verified from Synopsys reports).

*2) Case 2: Load capacitance is also significant*

The load capacitance consists the fan-outs from the neuron (same for both the Dig-N and MS-N) and the number of multipliers ($n$) connected to the output node (considered only for MS-N). Since the fan-out can be taken care of by inserting properly sized buffers for both Dig-N and MS-N, we only consider the effect of $n$ in this analysis and write the ratio $\left(\frac{P_{Dig}}{P_{MS}}\right)$ as shown in Eq. 25.

$$\frac{P_{Dig}}{P_{MS}} = \frac{630.6}{C_{eff}} = \frac{630.6}{n \times NC_{unit}}$$
$$= 0.75 \times \alpha \times \frac{N_{Dig}}{n \times N_{Ana,regular}} \times \frac{N-1}{N} \times L \quad (25)$$

Thus the worst-case energy benefit reduces by a factor of 10 when $n = 10$. This means that the MS-N is largely beneficial for a CNN where the number of connections to a neuron is limited. However, cascode topologies are shown to be useful when number of input/outputs are large in an analog design [25] and such structures can be employed for implementing a fully connected network using the proposed MS-N.

In summary, this analysis shows that the fundamental energy-benefit for the MS-N is a direct effect of *1) lower number of transistors* due to the ability to represent complex functions with intrinsic dynamics, *2) irregular slice structure, which trades off with the linearity of the neuron*. The leakage (at low frequency) and DVFS (at high frequency) in Dig-N further improves this energy-benefit at frequencies other than the near-threshold point (10 MHz) as the power consumption for MS-N is linear with frequency. The other significant trade-offs are between *total integrated noise* (which reduces by increasing device dimensions according to Eq. 11) *vs. bandwidth* (which degrades as per Eq. 6 when device dimensions are increased). The *DC offset due to device mismatch* can also be reduced at the cost of reduced *bandwidth*. The proposed design in section III alleviates these trade-offs by extending the bandwidth of the circuit using pole-zero compensation, which allows for increased device dimensions to reduce the effects of noise and mismatch.

## VI. EXPLOITING NEUROMORPHIC ERROR-RESILIENCY AT THE SYSTEM-LEVEL FOR LENET AND ALEXNET

To analyze the energy benefits and performance of the proposed MS-N, a cohesive circuit-algorithmic framework is developed that uses two well-known benchmark image recognition datasets, MNIST [15] and CIFAR-10 [16]. MNIST is a standard dataset of handwritten digits that contains 60,000 training and 10,000 test patterns of 28×28 pixel sized greyscale images of the digits 0-9. CIFAR10 is a more complex dataset that consists of 60,000 colored images belonging to 10 classes. Each image has 32×32 pixels. The first 50,000 images were used for for training and the last 10,000 images were used for testing.

Table III shows the different deep learning and fully connected implementations used to evaluate the datasets. It is to be noted that the learning architectures employed are the standard networks that have shown reasonable accuracy on the various benchmarks for low to medium-complexity applications [12][20], as sensor nodes targeted towards IoT and healthcare do not require deeper networks like ResNet or GoogLeNet to be implemented on the edge device. Each of the architectures shown in Table III was implemented using the widely used MatConvNet [21] platform, a deep learning toolbox used for training and evaluating the performance of the benchmark applications. While training, 16-bit precision was used to get a reasonable accuracy for the baseline network. However, for most ANNs, the bit precision can be scaled down to 8-bits without incurring any accuracy degradation as shown in Fig. 4 (error vs bit precision figure). Scaling below 8-bits may cause accuracy loss, a part of which can be restored by retraining the network. So, for lower bit precision (starting from 6 bits, down to 3 bits), incremental retraining was performed with bit width restrictions in place on the weights and neuron outputs to reclaim a significant portion of the accuracy ceded by scaling. Bit width scaling (and retraining) helps to get an optimized CMOS digital framework for our precision-constrained MS-multipliers. It also helps in obtaining an optimized digital baseline framework for fare energy/performance comparison with our MS-N. Hence, the software baseline implementation was aggressively optimized for performance.

TABLE III
LEARNING ANN ARCHITECTURES FOR EVALUATION

| Application | Learning Architecture |
| --- | --- |
| MNIST_FCN | 784×100×50×10 (2Hidden Layers) |
| MNIST_CNN | LeNet [12] |
| CIFAR_CNN | AlexNet [20] |

The trained baseline network with appropriate bit restrictions on the learnt weights is then evaluated on the testing set of the benchmark to obtain the performance or accuracy. The analog noise and mismatch models obtained from circuit simulations is incorporated in the software during evaluation phase. Since both DC offset and analog noise comes from the multiplier units that perform the multiplication of the weight values with the corresponding input, they are included within a modified weight value. The output at a particular neuron without noise and mismatch is given by Eq. 4, while the same output is calculated using Eq. 26 in presence of noise and mismatch.

$$V_{out} = F\left[A_v \times \sum_{k=1}^{n} w_k \left(v_k + \sqrt{A} + \Delta_k\right)\right]$$
$$= F\left[A_v \times \sum_{k=1}^{n} w_k \left(1 + \frac{\sqrt{A} + \Delta_k}{v_k}\right) v_k\right] \quad (26)$$

$A$ is the integrated noise power ($V^2$) within the bandwidth at a given bias current, while $\Delta_k$ is the DC offset in the $k$-th synapse. If the swing of $v_k$ is high, the effect of noise and mismatch is minimal. Fig. 23 presents the classification error for (a) MNIST_FCN, (b) MNIST_CNN and (c) CIFAR_CNN as a function of the non-ideality percentage (NIP), which we define as $NIP = \frac{\sqrt{A} + \Delta_k}{v_k} \times 100$. We observe that the worst case increase in error from the baseline is only 5.2% in 3-bit and 2.1% in the 8-bit proposed MS-N (both for CIFAR CNN) with a differential input swing of 400 mV and NIP of 1% ≈ 4 mV.

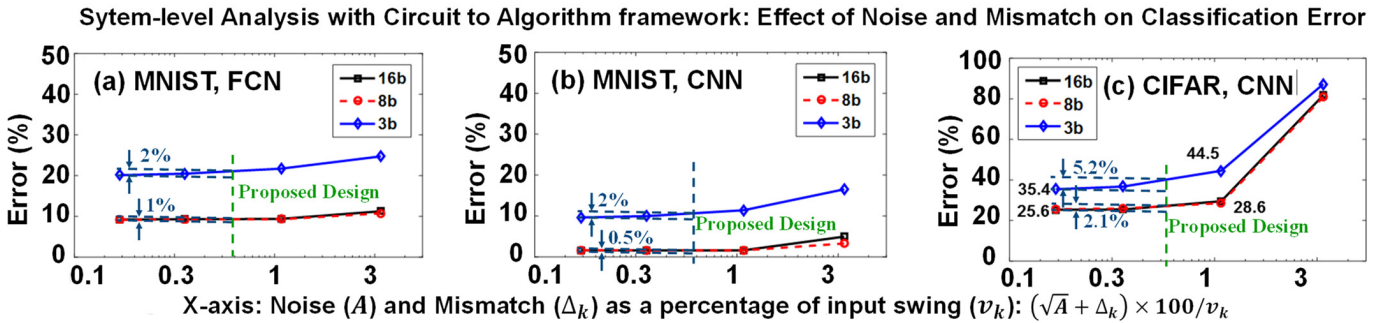

Fig. 23. System level simulation results for 3-bit, 8-bit and 16-bit MS-N: (a) MNIST [9] with FCN, (b) MNIST with CNN and (c) CIFAR-10 [10] with CNN

10TABLE IV
COMPARISON RESULTS WITH STATE-OF-THE-ART NEURON ARCHITECTURES

| Specification | SpiNNekar[1] [5] | TrueNorth [6] | Neurogrid [10] | BrainScaleS [9] | VMM [12] | Traditional MS-MAC (Sec II A.)[2] | Proposed Work (Section III)[3] |
|---|---|---|---|---|---|---|---|
| Meas./Sim. | Measured | Measured | Measured | Measured | Measured | Simulation | Analysis/Simulation |
| Neuron Type | Digital | Digital | Mixed-Signal | Mixed-Signal | Mixed-Signal | Mixed-Signal | Mixed-Signal |
| Applicable for | SNN | SNN | SNN | SNN | CNN | CNN | CNN |
| Architecture | ARM-core | Point neuron | Current mode | Current mode | Current mode | Current mode | Small signal |
| MAC based | No | No | No | No | Yes | Yes | Yes |
| Technology | 130 nm CMOS | 28 nm CMOS | 180 nm CMOS | 180 nm CMOS | 500 nm CMOS | 65 nm CMOS | 65 nm CMOS |
| Power/synapse | - | 254 nW | 390 pW | 19.5 µW | 5.310 µW | 0.9 µW | 10 nW |
| Frequency | 200 MHz | 1 kHz | 1 kHz | 125 MHz | 10 MHz | 1 MHz | 100 MHz |
| Energy/synapse | - | 254 fJ | 390 fJ | 156 fJ | 531 fJ | 900 fJ | 100 aJ |

[1]: Neurons are simulated on ARM-core (no physical implementation), [2,3]: The ANN architectures are not available on hardware

## VII. PERFORMANCE COMPARISON AND DISCUSSION

Table IV compares the performance of the proposed design with the existing neuron architectures. The proposed design achieves the best energy efficiency and can work at high frequencies which makes it suitable for neuromorphic computing applications. Though the power consumption per synapse is lower for Neurogrid, we must note that Neurogrid runs at a much lower frequency. The bias currents in the proposed design can be reduced for low frequency applications and have a better power per synapse value. Since the power in MS-N scales linearly with frequency, this will not degrade the energy efficiency.

It must be noted that the proposed work is based on simulation and analysis, while the other works presented in Table IV have measured data which considers the energy and latency of communication, memory fetch, data management and streaming which often proves to be a worse energy bottleneck than computation. However, in-memory [26] or near-memory [27] architectures help in reducing the memory-fetch power. As shown in [11], several power-reduction strategies such as event-driven computing, overlapping dendritic trees, island formation, hierarchical axonal structures, power-gating, multiplexed signaling and coordinated processing can be employed to reduce the communication energy, further exploiting the improved energy-efficiency of the proposed neuron at a system level. To account for the increased loading at the output nodes in a fully-functional ANN, a cascode topology as presented in [25] can be adopted as well. The proposed neuron model can thus be utilized to improve the power vs. frequency performance for the architectures demonstrated in the references shown in Table IV, with similar hardware for communication, memory fetch, data management and streaming.

## VIII. CONCLUSION

We have presented a MAC-based mixed-signal neuron architecture that can achieve extreme energy efficiency by employing a small signal voltage mode multiplication using a differential amplifier with resistive feedback. Compared to a traditional Dig-N, the proposed MS-N is ~1000X more energy-efficient at both low frequencies (<1 MHz) and very high frequencies (>500 MHz), and >100X more energy-efficient across frequencies in the range of 1-500 MHz without significantly affecting the classification error rate for digit/image recognition applications. Digital implementations can be duty cycled (using power gating) to reduce the power consumption. However, duty cycling does not reduce the on-time-energy-consumption. Moreover, for applications with low-frequency input signal, duty-cycled digital implementations require input and output FIFOs and suffer from the trade-off between FIFO size and frequency of turn-on/turn-off of the computation unit. The sources of the energy benefit in the MS-MAC for such scenarios is thoroughly analyzed and the significant trade-offs are identified, which are 1) energy vs. linearity, 2) energy vs total integrated noise and 3) energy vs. variability. A bandwidth extension technique is proposed which helps in alleviating the trade-offs and leads to a 0.1 fJ/MAC implementation of the 8-bit MS-N, which provides enough headroom for mimicking a biological neuron (20 fJ/MAC [11]) at a system level. As a future extension of this work, the memory fetch and communication energies of FCN and CNNs will be analyzed and implemented with near-memory computation, which promises to achieve an energy-efficient mixed-signal neural network.

APPENDIX A: FUNDAMENTAL ENERGY DISSIPATION AT ISO-FREQUENCY: 1-BIT DIGITAL VS. ANALOG

There has been a great deal of debate in the last few decades on the use of Analog/Mixed-signal computational units in neural networks. Digital had always been the preferred choice for Von-Neumann architectures as the scalability, reliability and accuracy is of prime importance in Von-Neumann. Neural networks, on the other hand, can tolerate the loss of accuracy to some extent and hence analog implementations can also be a viable option. Even though it is usually believed that analog consumes less energy, it is not true that analog is always better than digital. Fig. 24 presents two simple implementations - one CMOS digital inverter and one analog amplifier. Assuming that the inverter is dynamic power dominated and the amplifier operates in sub-threshold, we can write the equations for power consumption for the two cases.

$$P_{Dig} = C_L V_{DD,Dig}^2 f \quad (27)$$

And
$$P_{Ana} = V_{DD,Ana} I$$
$$= V_{DD,Ana} \times (2\pi \times \eta V_T \times C_L \times BW) \quad (28)$$

Where the terms have their usual meaning as shown previously in Eq. 1 and Eq. 6. For iso-frequency ($\approx$ 100 MHz, where inverter is dynamic power dominated), $f = BW$, and the ratio $\left(\frac{P_{Dig}}{P_{Ana}}\right)$ can be written as shown in Eq. 29.

$$\frac{P_{Dig}}{P_{Ana}} = \frac{V_{DD,Dig}^2}{V_{DD,Ana} \times (2\pi \times \eta V_T)} \quad (29)$$

From the theory of CMOS inverters, the minimum supply voltage $V_{DD,Dig}$ can be calculated from Eq. 30 for a given maximum frequency ($f$)

$$f = \frac{1}{t_r + t_f} = \frac{1}{2.2 R_p C_L + 2.2 R_n C_L}$$
$$= \frac{\beta_p (V_{DD,Dig} - |V_{T,p}|) \beta_n (V_{DD,Dig} - V_{T,n})}{2.2 C_L [\beta_p (V_{DD,Dig} - |V_{T,p}|) + \beta_n (V_{DD,Dig} - V_{T,n})]} \quad (30)$$

For $f \approx$ 100 MHz, $V_{DD,Dig}$ is calculated as 0.35 V in 65 nm CMOS technology (very close to the results shown in Fig. 5). Assuming $V_{DD,Dig} = 0.4$ V and $V_{DD,Ana} = 1$ V, $\left(\frac{P_{Dig}}{P_{Ana}}\right)$ Eq. 31 can be obtained from Eq. 29.

$$\frac{P_{Dig}}{P_{Ana}} = \frac{(0.4)^2}{1 \times (2\pi \times 1.3 \times 0.026)} = 0.75 \quad (31)$$

Which is the same factor of 0.75 in Eq. 24. If $V_{DD,Ana}= 0.75$ V, $\left(\frac{P_{Dig}}{P_{Ana}}\right) = 1$, which is an interesting result, because it means that to achieve a certain bandwidth, both digital and analog fundamentally consumes the same amount of power. However, this is because the designs considered in this example are relatively simple, and both digital and analog consists exactly 2 transistors. In a larger design, the $\alpha \times \frac{N_{Dig}}{N_{Ana}}$ factor will come into the picture as shown in Eq. 24. This accounts for the large number of transistors in a complex digital design where the intrinsic dynamics in analog allows having lower number of transistors. The irregular slices in the proposed design further improves the energy-efficiency at the cost of linearity, which is again possible only in analog. Hence, it is not the analog alone, but the combination of approximate computing, intrinsic dynamics and design trade-offs that triumphs over digital implementations. In scenarios where those trade-offs are not acceptable, digital is and will still have the upper hand. However, neuromorphic computing provides a platform where such trade-offs are not only tolerated, but can also be utilized, which makes the design of such networks a truly interesting problem.

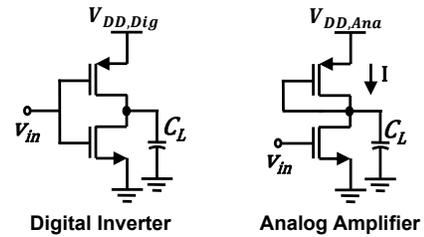

Fig. 24. Digital Inverter vs. Analog Amplifier: simple implementations